\documentclass[12pt]{article}
\usepackage{graphicx}
\usepackage{amssymb}
\usepackage{amsmath}

\newcommand{\bear}{\begin{array}}  
\newcommand {\eear}{\end{array}}
\newcommand{\bea}{\begin{eqnarray}}   
\newcommand{\eea}{\end{eqnarray}}
\newcommand{\beq}{\begin{equation}}   
\newcommand{\eeq}{\end{equation}}
\newcommand{\bef}{\begin{figure}}  \newcommand 
{\eef}{\end{figure}}
\newcommand{\bec}{\begin{center}}  \newcommand 
{\eec}{\end{center}}

\begin{document}

\begin{titlepage}

\begin{flushright}
IPMU10-0191 \\
\end{flushright}

\vskip 1.35cm

%\preprint{ICRR-Report-???}
%\preprint{IPMU 09-????}

\begin{center}

{\large \bf
Charge-breaking constraints on left-right mixing of stau's 
}

\vskip 1.2cm

Junji Hisano$^{a,b}$ 
and Shohei Sugiyama$^{a,c}$

\vskip 0.4cm

{ \it $^a$Department of Physics, Nagoya University, Nagoya 464-8602,
 Japan}\\
{\it $^b$Institute for the Physics and Mathematics of the Universe,
University of Tokyo, Kashiwa 277-8568, Japan}\\
{\it $^c$Department of Physics, University of Tokyo, Tokyo 113-0033,
 Japan }

%\date{\today}

\begin{abstract} 
  In the minimal supersymmetric standard model, large left-right
    mixing of stau's is sometimes intriguing from phenomenological
    viewpoints. However, too large left-right mixing is not acceptable
    since the electroweak-breaking vacuum becomes metastable.  In this
    paper the vacuum transition rate is evaluated by using
    semiclassical techniques, and constraints on parameters of the
    model are shown.  In the calculation the bounce solution is derived
    in the multifield space.  These constraints are also applied to the
    case of the low-energy minimal gauge mediation model.
\end{abstract}

%\pacs{95.35.+d}

%\maketitle

\end{center}
\end{titlepage}

%%%%%%%%%%%%%%%%%%%%%%%%%%%%%%%%%%%
\section{Introduction}\label{sec:intro}
%%%%%%%%%%%%%%%%%%%%%%%%%%%%%%%%%%%
The minimal supersymmetric standard model (MSSM) has two Higgs
doublets, and the ratio of their vacuum expectation values,
$\tan\beta(\equiv \langle H_u\rangle/ \langle H_d\rangle$), is an
important parameter when the phenomenology is discussed. It is
expected to be between $\sim 2$ and $\sim 60$ if the Yukawa coupling
constants for the third generation fermions are perturbative below the
GUT scale.

 Large $\tan\beta$ is favored from several phenomenological
  viewpoints. First, the light Higgs boson mass depends on $\tan\beta$
  with a high mass at larger $\tan\beta$. The muon $(g-2)$ anomaly
  also favors large $\tan\beta$ since the SUSY correction is
  proportional to $\tan\beta$ \cite{Czarnecki:1998nd}.  Second, the
  Yukawa unification of the third generation fermions in the SO(10)
  SUSY GUT requires large $\tan\beta$ \cite{Ananthanarayan:1991xp}.

   Third, the SUSY CP problem is automatically solved when $A$ and
    $B_{\mu}$ parameters vanish at high energy scale.  The $A$ and
    $B_{\mu}$ parameters are for trilinear and Higgs bilinear
    soft-SUSY breaking couplings, respectively. In the case, large
  $\tan\beta$ is predicted, because $\tan\beta$ is proportional to
  inverse of the $B_{\mu}$ parameter, while the $B_{\mu}$ parameter is
  only radiatively generated. It is pointed out in
  Refs.~\cite{Dine:1996xk,Rattazzi:1996fb} that in the 
   low-energy minimal gauge mediation (MGM) model, in which the messenger scale is
  around $10^{(5-6)}$~GeV, $\tan\beta$ is predicted to be $(50-60)$.
  The radiative correction induces tiny $B_\mu$ parameter due to the low
  messenger scale.

Forth, in some gauge mediation models, stau is the next-lightest SUSY
particle (NLSP). It is so long-loved that the big bang nucleosynthesis
(BBN) may be destroyed \cite{Pospelov:2006sc}. When the left-right
mixing term in the stau mass matrix, which is proportional to
$\tan\beta$, is large, the annihilation of stau pair is enhanced due
to the Higgs $s$-channel exchange so that the primordial abundance of
stau NLSP is reduced \cite{Ratz:2008qh}.

 While large $\tan\beta$ sometimes intriguing as mentioned above, it is
known that large $\tan\beta$ solutions may suffer from vacuum
instability \cite{Rattazzi:1996fb}. When the left-right mixing term
for stau's is increased, electric charge-breaking minimum in the
scalar potential appears and it becomes deeper than than the
``ordinary'' electroweak-breaking minimum. The lifetime of the
electroweak-breaking vacuum is required to be longer than the age of
the universe.

In this paper, we derive upperbound on the left-right mixing term for
stau's as a function of the left- and right-handed slepton masses, by
imposing that the electroweak-breaking vacuum has longer lifetime than
the age of the universe. The corresponding lowerbound on the
stau mass is also shown as a function of the left- and right-handed slepton masses.  It is found that the constraints are
insensitive to the parameters in the Higgs potential, including
$\tan\beta$ itself. When the LHC experiment discovers stau's or gives
lowerbound on the mass, the stability of the electroweak-breaking
vacuum would be useful to derive the parameters in the stau mass matrix.

The quantum transition rate of the metastable vacuum is estimated by
semiclassical technique \cite{Coleman:1977py}. We evaluate numerically
bounce configuration in three-fields space (left- and right-handed
stau's and Higgs boson). The quantum transition rate is evaluated in
the previous works, though the bounce configuration is approximated to
be one-dimensional \cite{Rattazzi:1996fb,Ratz:2008qh}.  Thus, our 
result is more accurate than those.

This paper is organized as follows. In next section, the quantum
transition rate of the electroweak-breaking vacuum to the true one is
evaluated, and stability of the electroweak-breaking vacuum is
discussed. The constraints on parameters for stau mass matrix are
derived there. In Section 3 the constraints on the low-energy 
  MGM model are shown.  Section 4 is devoted to conclusions and
discussion.

%%%%%%%%%%%%%%%%%%%%%%%%%%%%%%%%%%%
\section{Stability of  vacuum}
%%%%%%%%%%%%%%%%%%%%%%%%%%%%%%%%%%%

Stability of the vacuum in the MSSM in the case that $\tan\beta$ is
large have been studied in Ref.~\cite{Rattazzi:1996fb}.  In this
section we follow their argument and give a more reliable evaluation for
vacuum stability.

In gauge mediation models and the constrained MSSM which demands
unification of soft scalar squared masses at high scale,
sleptons are lighter than squarks because in the former cases slepton masses
are proportional to small squared gauge couplings $g_Y^2$ of $U(1)_Y$
and $g_2^2$ of $SU(2)_L$ and in the latter case squarks receive large
gluino contribution through renormalization group (RG) running.
If $\tan\beta$ is large the lightest sfermion is stau for large
left-right mixing, which is proportional to the
Yukawa coupling of the tau lepton,
\begin{equation}
 y_{\tau}=\frac{m_{\tau}}{v_d}\sim\frac{\tan\beta}{100}.
\end{equation}
Too large $\tan\beta$ makes stau tachyonic and classical
stability of the electroweak-breaking vacuum is lost. Although the electroweak-breaking vacuum is
classically stable, it may be unstable by quantum tunneling effect. 
If the electroweak-breaking vacuum is not a global minimum but local one
in the potential, it  would collapse into the global one eventually.
Such a metastable vacuum is viable only when its lifetime is longer than the
age of the universe.

To see the situation described above in detail we write the 
scalar potential for neutral component of up-type Higgs $H_u$,
left-handed stau $\tilde{L}$ 
and right-handed stau $\tilde{\tau}_R$ as follows,
\begin{align}\label{RS}
\begin{split}
  V=&\:(m_{H_u}^2 + \mu^2) |H_u|^2 + m_{\tilde{L}}^2 |\tilde{L}|^2 +
  m_{\tilde{\tau}_R}^2 |\tilde{\tau}_R|^2 - (y_{\tau}\mu H_u^\star
 \tilde{L} \tilde{\tau}_R + \text{h.c.})
  + y_{\tau}^2 |\tilde{L} \tilde{\tau}_R|^2  \\
 &+ \frac{g_2^2}{8}(|\tilde{L}|^2
  + |H_u|^2)^2 + \frac{g_Y^2}{8}(|\tilde{L}|^2 -2 |\tilde{\tau}_R|^2 -
  |H_u|^2)^2 + \frac{g_2^2 + g_Y^2}{8} \delta_H |H_u|^4,
\end{split}
\end{align}
where $\mu$ is supersymmetric Higgs mass and $m^2$'s are soft squared
masses of each scalar. For simplicity, $\mu$ is assumed to be real.
Terms including down-type Higgs $H_d$ are ignored since the vacuum expectation value (VEV)
is small for large $\tan\beta$. The scalar potential includes an
RG-improved term, which is the last term in Eq.~(\ref{RS}) and
reflects quartic Higgs interaction induced by a loop diagram of the top
quark:
\begin{equation}
 \delta_H = \frac{3}{\pi^2}\frac{y_t^4}{g_Y^2+g_2^2}\log\frac{m_{\tilde{t}}}{m_t}.
\end{equation}
Here, $y_{t}$ is the Yukawa coupling of the top quark and $m_{t}$ and
$m_{\tilde{t}}$ are the mass of the top and stop, respectively.
Typical value of $\delta_H$ is $\sim 1$.
In addition, loop diagrams of the stop also induces the quartic
interaction and may make sizable contribution in some cases.
It is proportional to $A_t^4$, where $A_t$ is an $A$ parameter for stop interaction.

 The $H_u$ mass term, $m^2_{H_u}+\mu^2$, is negative so that
$H_u$ gets a VEV $v_u$.  Expanding around this electroweak-breaking
vacuum, we have
\begin{align}
\begin{split}\label{potential}
  V=&\: m_{\phi}^2 \phi^2 
 +\left(m_{\tilde{L}}^2 + \frac{g_2^2-g_Y^2}{4} v_u^2\right)\tilde{L}^2 
 + \left(m_{\tilde{\tau}_R}^2 + \frac{g_Y^2}{2}v_u^2\right) \tilde{\tau}_R^2 
 - 2 y_{\tau} \mu v_u\tilde{L}\tilde{\tau}_R \\
&- 2 y_{\tau} \mu \phi\tilde{L}\tilde{\tau}_R
 + \frac{g_2^2-g_Y^2}{2}v_u \phi\tilde{L}^2 
 + g_Y^2 v_u \phi \tilde{\tau}_R^2
 + \frac{m_{\phi}^2}{v_u}\phi^3 + \cdots,
%  + \frac{m_{\phi}^2}{4v_u^2}\phi^4 
%  + \frac{g_Y^2+g_2^2}{8}\tilde{L}^4 \\
% & + \frac{g_Y^2}{2}\tilde{\tau}_R^4 
%  + \frac{g_2^2-g_Y^2}{4}\phi^2 \tilde{L}^2
%  + \frac{g_Y^2}{2} \phi^2 \tilde{\tau}_R^2
%  + \left(y_{\tau}^2 - \frac{g_Y^2}{2}\right)\tilde{L}^2\tilde{\tau}_R^2,
\end{split}
\end{align}
where $H_u=v_u+\phi$ and
$m_{\phi}^2=\frac{g_Y^2+g_2^2}{2}(1+\delta_H)v_u^2=(1+\delta_H)\sin^2\beta\,
m_Z^2$. Ellipsis stands for quartic terms.  We show only terms with
the real parts of scalar bosons in above scalar potential. The first
line of Eq.~(\ref{potential}) gives a squared mass matrix of stau's,
\begin{equation}
 \mathcal{M}_{\tilde{\tau}} ^2 = \begin{pmatrix}
 m^2_{\tilde{L}}+(\frac{1}{2}-s_W^2) m_Z^2 & \mu y_{\tau} v_u \\
 \mu y_{\tau} v_u & m^2_{\tilde{\tau}_R} +s_W^2 m_Z^2
				 \end{pmatrix}.
\end{equation}
Here $s_W^2=g_Y^2/(g_Y^2+g_2^2)$ and we take a large $\tan\beta$ limit.
Classical stability of the electroweak-breaking vacuum is equivalent to the
positivity of the smaller eigenvalue of the matrix
$\mathcal{M}^2_{\tilde{\tau}}$. 

Even if the classical stability condition is satisfied, the first term
in the second line of Eq.~(\ref{potential}) could generate a global
minimum where $\langle\phi\rangle, \langle \tilde{L}\rangle$, and
  $\langle\tilde{\tau}_R\rangle\neq 0$ and make the
electroweak-breaking vacuum metastable. 

 Quantum transition rate of the metastable vacuum is estimated by
semiclassical technique \cite{Coleman:1977py}.  In this technique the
imaginary part of energy of the false vacuum, which is proportional to
the transition rate is evaluated using path integral method in
Euclidean spacetime.  The path integral is dominated by so-called
bounce configuration, $\varphi_i=\bar{\varphi}_i(t,\vec{x})$, where
$\varphi_1=\phi,~\varphi_2=\tilde{L}$ and $\varphi_3=\tilde{\tau}_R$
in our case.  It is a stationary point of the action and satisfies
boundary conditions
$\lim_{t\rightarrow\pm\infty}\bar{\varphi}_i(t,\vec{x})=\varphi_i^f$,
where $\varphi_i^f$ are values of the fields at false vacuum.  It is
known that we can take an $O(4)$ symmetric solution
\cite{Coleman:1977th}.  The $O(4)$ symmetric Euclidean action is
\begin{equation}
 S_E[\varphi(r)]=2\pi^2\int dr r^3\left[\sum_{i=1}^3\left(
\frac{d\varphi_i}{dr}\right)^2 + V(\varphi)\right],
\end{equation}
where $r$ is a radial coordinate in four-dimensional spacetime.
The equation of motion and boundary conditions are 
\begin{gather}
  2\frac{d^2\varphi_i}{d r^2}
+\frac{6}{r}\frac{d\varphi_i}{dr}
=\frac{\partial V}{\partial\varphi_i}(\varphi)
,\\
 \lim_{r\rightarrow\infty}\bar{\varphi}_i(r)=\varphi_i^f,\qquad
  \frac{d\bar{\varphi}_i}{dr}(0)=0.
\end{gather}
Finally the vacuum transition rate per unit volume is evaluated as follows,
\begin{equation}
 \Gamma/V=A e^{-B}.
\end{equation}
The prefactor $A$ is the fourth power of the typical scale in the
potential. Its precise value is hard to calculate but the transition rate is
not sensitive to it. On the other hand, $B$ has an importance in the
evaluation and
\begin{equation}
 B=S_E[\bar{\varphi}(r)]-S_E[\varphi^f].
\end{equation}
If we demand that $\Gamma/V$ is much smaller than the fourth power of
the present 
Hubble expansion rate $H_0=1.5\times 10^{-42}$~GeV and assume $A$ is
$(100~{\rm GeV})^4$, we have constraint on metastability that 
$B\gtrsim 400$.

For numerical calculation of bounce solutions we used the method of
Ref.~\cite{Konstandin:2006nd}. Search for bounce solution with single
field is easily achieved by overshooting/undershooting method in which
one scans initial values $\varphi(0)$ around the global minimum. This
scan is difficult in the case of multifield. The authors of
Ref.~\cite{Konstandin:2006nd} found that in one-dimensional space the
initial value $\varphi(0)$ is obtained by using a modified
potential. Once we have a bounce solution in one-dimensional space,
gradually increasing the spacetime dimension we obtain the
four-dimensional solution. Actual calculation is based on a
discretized equation of motion.

%%%%%%%%%%%%%%%%%%FIGURE%%%%%%%%%%%%%%%%%%%

\begin{figure}[t]
 \begin{center}
   \includegraphics[width=0.7\linewidth]{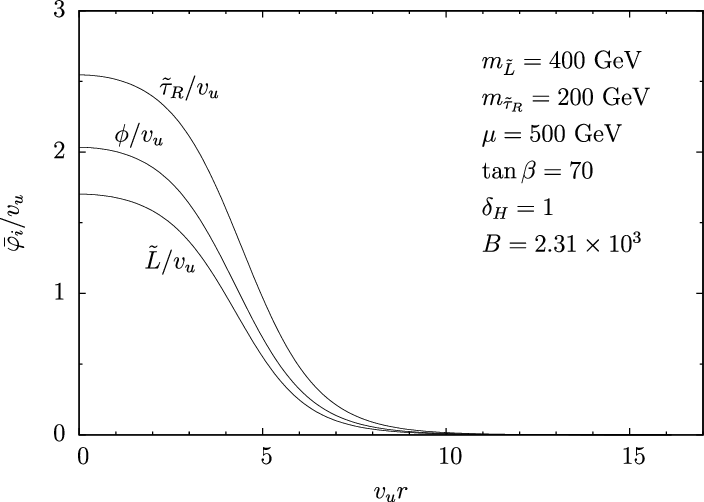}
   \caption{Bounce configuration $\bar{\varphi}_i(r)$ as a function of
  the radial coordinate $r$.
 Parameters of potential and $B$ are shown in figure.
}
   \label{fig:bounce}
 \end{center}
\end{figure}

%%%%%%%%%%%%%%%%%%%%%%%%%%%%%%%%%%%%%%%%%

This method is unsuitable yet to the case that the global
minimum is much deeper than local one. Because in that case the
modified potential is quite different from original one so that it is
difficult to obtain the one-dimensional solution.  For such parameter
regions, starting from the solution obtained in region where two
minima are nearly degenerate, we gradually change the parameters of
the potential and reach the solution.
Fig.~\ref{fig:bounce} is a bounce solution obtained by this method.  
% A trajectory is slightly different from a straight line assumed in
% Ref.~\cite{Rattazzi:1996fb}.  {When $\phi\simeq0.2v_u$,} $\tilde{L}$
% and $\tilde{\tau}_R$ approach almost 0.

%%%%%%%%%%%%%%%%%%FIGURE%%%%%%%%%%%%%%%%%%%

\begin{figure}[h]
\begin{center}
    \includegraphics[width=0.7\linewidth]{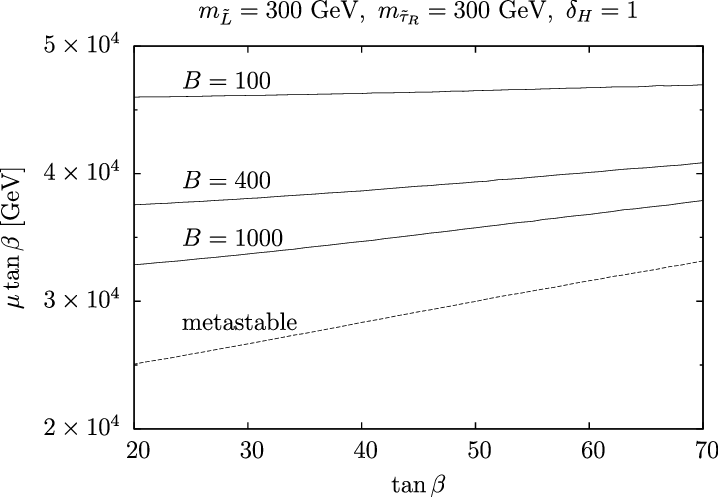}\\
 \vspace{1.5cm}
    \includegraphics[width=0.7\linewidth]{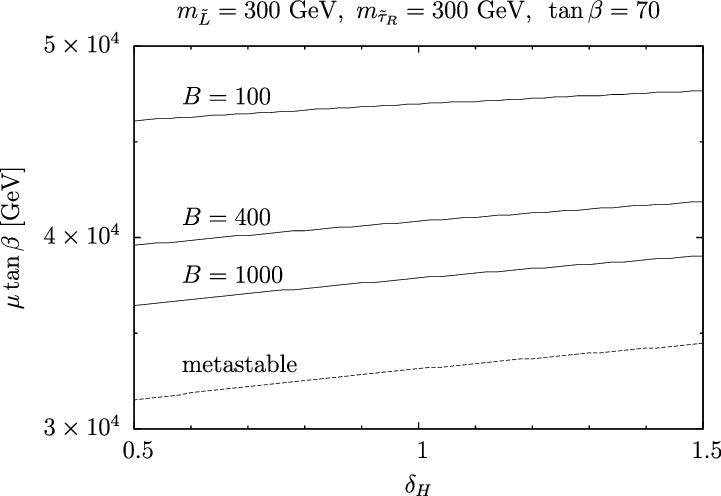}
\end{center}   
\caption{ Solid lines are contours of $B=100,~400$ and $1000$ in
 $\tan\beta$--$\mu\tan\beta$ plane 
 (top) and $\delta_H$--$\mu\tan\beta$ plane (bottom).
 The electroweak-breaking vacuum becomes metastable above broken lines.
  }
   \label{fig:irrelevant}
\end{figure}

%%%%%%%%%%%%%%%%%%%%%%%%%%%%%%%%%%%%%%%%%

%%%%%%%%%%%%%%%%%%FIGURE%%%%%%%%%%%%%%%%%%%

\begin{figure}[t]
 \begin{center}
   \includegraphics[width=0.7\linewidth]{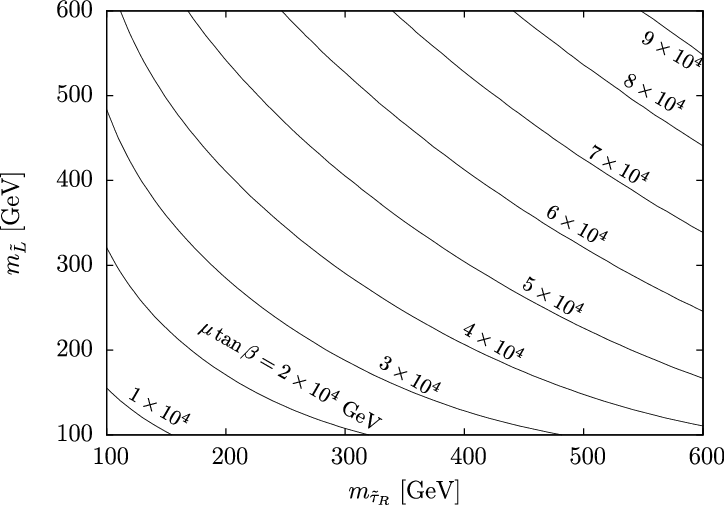}
   \caption{Contour plots for upperbound on $\mu\tan\beta$ that 
satisfies $B\ge 400$ in $m_{\tilde{\tau}_R}$--$m_{\tilde{L}}$ plane.
}
   \label{fig:se400}
 \end{center}
\end{figure}

%%%%%%%%%%%%%%%%%%%%%%%%%%%%%%%%%%%%%%%%%

%%%%%%%%%%%%%%%%%%FIGURE%%%%%%%%%%%%%%%%%%%

\begin{figure}[t]
 \begin{center}
   \includegraphics[width=0.7\linewidth]{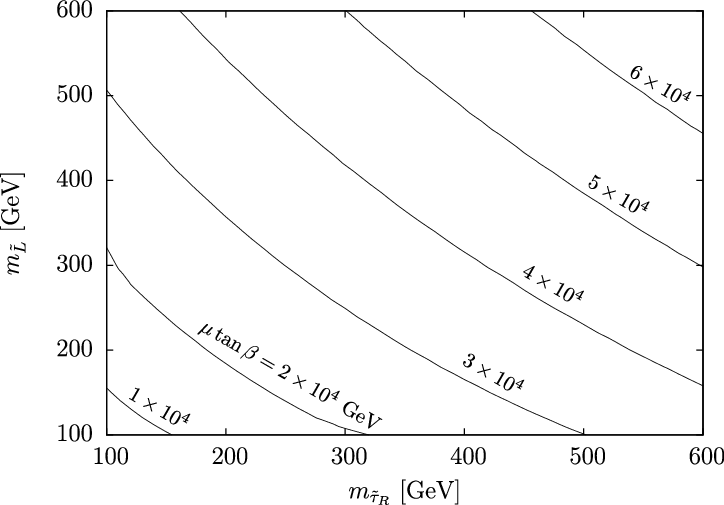} 
   \caption{Contour plots for upperbound on $\mu\tan\beta$ that
  satisfies condition that the electroweak-breaking vacuum is global
  minimum.
}
   \label{fig:metastable}
 \end{center}
\end{figure}

%%%%%%%%%%%%%%%%%%%%%%%%%%%%%%%%%%%%%%%%%

%%%%%%%%%%%%%%%%%%FIGURE%%%%%%%%%%%%%%%%%%%

\begin{figure}[t]
 \begin{center}
   \includegraphics[width=0.7\linewidth]{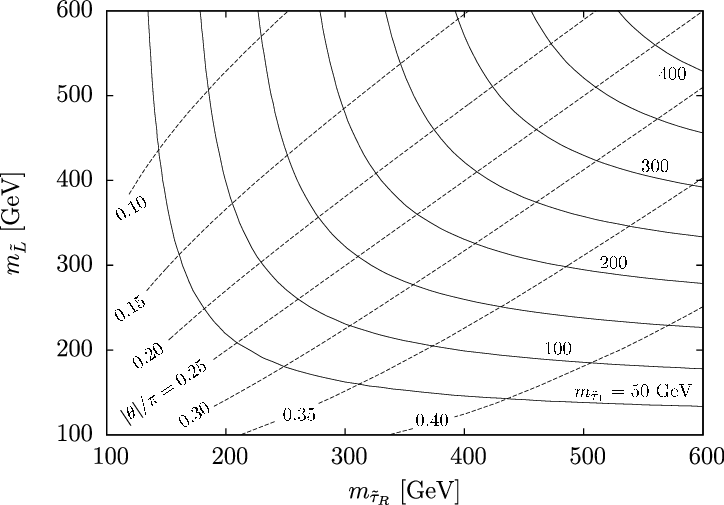}
   \caption{Contour plots for stau mass $m_{\tilde{\tau}_1}$
  (solid lines) and stau mixing angle $\theta_{\tilde{\tau}}$ (broken
  lines) in $m_{\tilde{\tau}_R}$--$m_{\tilde{L}}$ plane 
when $\mu\tan\beta$ is taken to be upperbound shown in Fig.~\ref{fig:se400}.
}
   \label{fig:mass}
 \end{center}
\end{figure}

%%%%%%%%%%%%%%%%%%%%%%%%%%%%%%%%%%%%%%%%%

For the potential of Eq.~(\ref{potential}) relevant parameters to vacuum stability are 
$m^2_{\tilde{L}},~m^2_{\tilde{\tau}_R},~\mu$, $\tan\beta$ and $\delta_H$.
Since the coefficient of quadratic term $\tilde{L}\tilde{\tau}_R$ is
proportional to $\mu\tan\beta$ and that of cubic term
$\phi\tilde{L}\tilde{\tau}_R$ is nearly proportional to $\mu\tan\beta$,
the bounce solution is sensitive to $\mu\tan\beta$, but not $\tan\beta$ itself.
To confirm this, we plot contours of $B$ changing $\mu\tan\beta$ and
$\tan\beta$ while other parameters are fixed
(Fig.~\ref{fig:irrelevant} (top)). Irrelevance of $\tan\beta$ is clearly
seen in the figure. Weak dependence on $\tan\beta$ is generated by the quartic term
$y_{\tau}^2\tilde{L}^2\tilde{\tau}_R^2$ in Eq.~(\ref{RS}). This term lifts
the global minimum and increases $B$ a little.  

Next, we also plot contours changing $\mu\tan\beta$ and $\delta_H$ in
Fig.~\ref{fig:irrelevant} (bottom).  
When $A_t$ is negligible, $\delta_H$ mainly depends on the
stop mass. If we take $m_{\tilde{t}}$ in the
range 600~GeV $\leq m_{\tilde{t}}\leq$ 2~TeV, $\delta_H$ has the value
from 0.65 to 1.3.  Larger value of $\delta_H$ stabilizes the
electroweak-breaking vacuum and slightly increases $B$.
Thus, it is found that the transition rate is sensitive to only 
$m^2_{\tilde{L}},~m^2_{\tilde{\tau}_R}$, and $\mu\tan\beta$ among parameters 
in the potential of Eq.~(\ref{potential}).

In Fig.~\ref{fig:se400}, we show upperbound on $\mu\tan\beta$ that
satisfies the metastability condition $B\ge 400$.  In numerical
  calculation we set $\delta_H=1$ and $\tan\beta=70$.  It is found
from Fig.~\ref{fig:irrelevant} (top) that dependence on $\tan\beta$
changes the result at most 5\%.  We also evaluate that the dependence
on $\delta_H$ is less than 1\% in the range that $0.65\leq\delta_H\leq
1.3$.  We also show upperbound on $\mu\tan\beta$ that satisfies a
rigorous stability condition  (the electroweak-breaking vacuum is global
  minimal in the potential) in Fig.~\ref{fig:metastable}.  The
metastability condition gives about $50\%$ looser constraint on
$\mu\tan\beta$ than that the rigorous stability condition.

Upperbound on $\mu\tan\beta$ gives lowerbound on stau mass. We plot
stau mass and mixing angle when $\mu\tan\beta$ is taken to be its
maximum value allowed by metastability in Fig.~\ref{fig:mass}.
Since LEP2 experiments give lowerbounds on stau mass as 
$m_{\tilde{\tau}_1}\gtrsim$ 80--90~GeV \cite{Nakamura:2010zz},
in these region in Fig.~\ref{fig:mass} the metastability condition gives
severer constraint on $\mu\tan\beta$ than the experimental bounds.

 We fit this result  under the metastability condition to a function
as follows,
\begin{equation}
  \mu\tan\beta <213.5 \sqrt{m_{\tilde{L}}m_{\tilde{\tau}_R}}
-17.0(m_{\tilde{L}}
+m_{\tilde{\tau}_R})
+4.52\times 10^{-2} {\rm ~GeV}^{-1}
(m_{\tilde{L}}-m_{\tilde{\tau}_R})^2
 - 1.30\times 10^4  {\rm ~GeV}.
\label{fitting}
\end{equation}
The difference of this fit is less than 1\% in the region where
$m_{\tilde{\tau}_1}\gtrsim$ 80~GeV.  This would be useful when
considering large left-right mixing of stau's.

%%%%%%%%%%%%%%%%%%%%%%%%%%%%%%%%%%%%%%
\section{Application} 
%%%%%%%%%%%%%%%%%%%%%%%%%%%%%%%%%%%%%%
As an application of the previous section, we consider the low-energy MGM
  model. The MGM model is the gauge mediation model with a constraint that the
$B_\mu$ term vanishes at the messenger scale. This model is a solution
for the SUSY CP problem in Refs.~\cite{Dine:1996xk,Rattazzi:1996fb}, as
mentioned in Introduction.

In this model $\tan\beta$ is an output of the model rather than input.
At the tree level, $B_{\mu}$ and $\tan\beta$ are related as follows,
\begin{equation}
 -\frac{2B_\mu}{m_{H_u}^2+m_{H_d}^2+2\mu^2}=\sin 2\beta\simeq
  \frac{2}{\tan\beta}. 
\end{equation}
Since $B_{\mu}$ is generated only through RG running, large 
$\tan\beta$ is predicted in the low-energy MGM model.

There are two classes of contributions to $B_{\mu}$. The first one
arises from one-loop diagrams of gaugino. The other class is generated by
$A$ terms. Since $A$ terms also vanish at the messenger scale, this
contribution arises at two loop level. The largest one among $A$ term
contributions is proportional to squared Yukawa coupling of the top (or bottom) quark
and squared $SU(3)_C$ gauge coupling. Because of these large couplings, 
the $A$ term contributions are comparable with the gaugino one despite the
extra loop. These two contributions have opposite signs, and $B_{\mu}$ is
smaller than naive expectation in the low-energy MGM model.

We assume that messengers are $\bf{5}+\bf{5}^*$ in $SU(5)$ and write
couplings to SUSY-breaking sector field $S$ as follows,
\begin{equation}
 W=\lambda_T S \bar{q}{q} + \lambda_D S \bar{l}{l},
\end{equation}
where, $q$ and $l$ are $SU(3)_C$ triplet and $SU(2)_L$ doublet in
messenger, respectively. $\lambda$'s are coupling constants and $S$ has VEVs
$\langle S \rangle=M+\theta^2 F_S$.
When $\lambda$'s are equal at the GUT scale, they differ at messenger
scale due to RG running, which takes $\lambda_T$ larger than $\lambda_D$ at
low energy. The gaugino masses are expressed as follows,
\begin{equation}
 M_i=\frac{\alpha_i}{4\pi}\frac{F_S}{M}g(x_i),
\end{equation} 
where $\alpha_i=g^2_i/4\pi$ and the loop function $g$ is 
\begin{equation}
 g(x)=\frac{1}{x^2}[(1+x)\log(1+x)+(1-x)\log(1-x)].
\end{equation}
$x_{2,3}$ are defined by $x_2=F_S/\lambda_D M^2$, $x_3=F_S/\lambda_T M^2$ and 
$g(x_1)=\frac{2}{5}g(x_3)+\frac{3}{5}g(x_2)$.  The difference in $\lambda$'s
affects the gaugino masses only through the loop function and its
effect becomes large in the low-energy MGM model where $F_S\sim M^2$.

\begin{figure}[t]
 \begin{center}
   \includegraphics[width=0.5\linewidth]{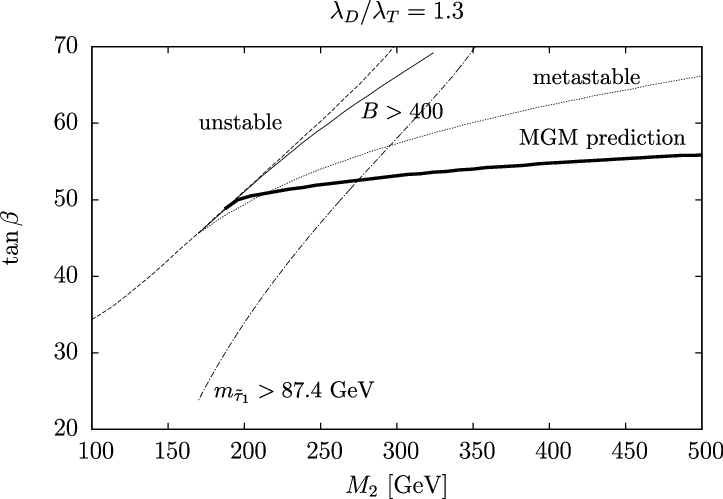}\\ [.5cm]
   \includegraphics[width=0.5\linewidth]{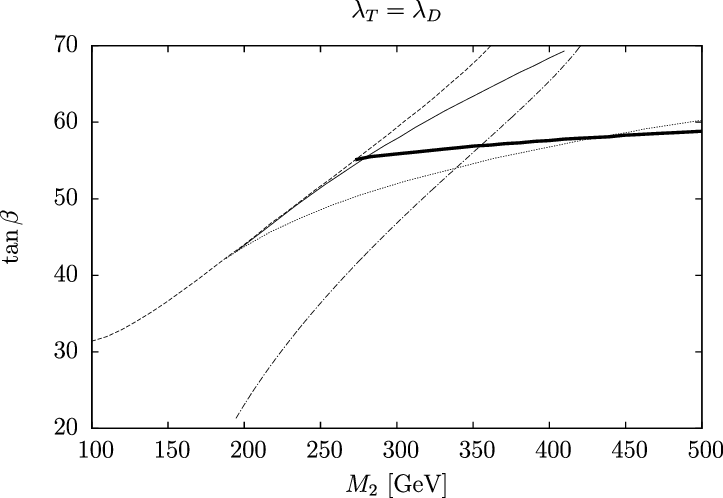}\\ [.5cm]
   \includegraphics[width=0.5\linewidth]{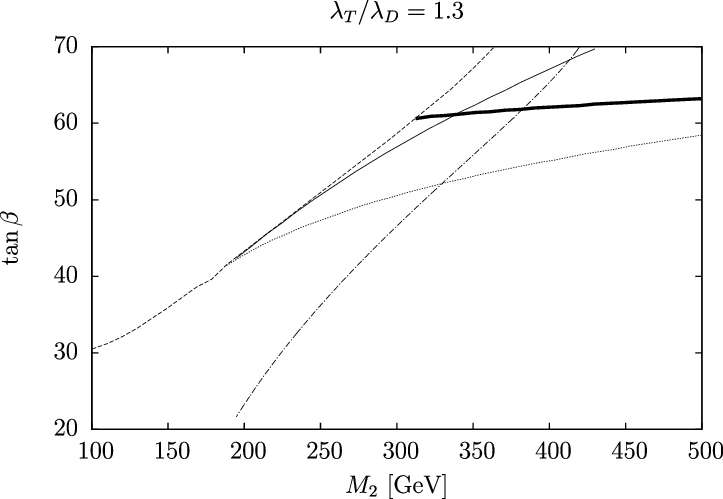} 
   \caption{(Meta-)stability constraints in the low-energy MGM
       model. Horizontal axis corresponds to gaugino mass of
     $SU(2)_L$, and vertical axis corresponds to $\tan\beta$. Heavy
     lines indicate predicted value of $\tan\beta$ in MGM.
     Electroweak-breaking vacuum is unstable above broken lines and
     metastable above dotted lines. Below solid lines metastability
     condition $B>400$ is satisfied. In the right-hand side of
     dot-dashed lines stau mass is larger than experimental
     lowerbound. We take $\lambda_D=1.3\lambda_T$ (top),
     $\lambda_T=\lambda_D$ (middle), $\lambda_T=1.3\lambda_D$ (bottom)
     and $\max(x_2,x_3)=0.95$ in all cases.  }
   \label{fig:mgm}
 \end{center}
\end{figure}

In Fig.~\ref{fig:mgm} we show the prediction of the low-energy MGM model and constraints by
(meta-)stability of the vacuum. 
We take $\lambda_D=1.3\lambda_T$ (top), 
$\lambda_T=\lambda_D$ (middle), $\lambda_T=1.3\lambda_D$ (bottom) 
and $\max(x_2,x_3)=0.95$. 
These results are obtained using SuSpect \cite{Djouadi:2002ze} with
modification for splitting in couplings of messenger.
In the top and bottom figures the ratio of loop
functions is $g(0.95):g(0.95/1.3)=1.14:1$.
In gauge mediation models current experimental lowerbound on stau mass is 87.4~GeV \cite{Nakamura:2010zz}.
% In the case of $\lambda_T=1.3\lambda_D$, $\tan\beta$ is larger than 60
% and the metastability condition gives severer constraint on lower limit
% of $M_2$ than experimental lowerbound on stau mass gives.
% This is explained as follows.
In Fig.~\ref{fig:mgm} the magnitude of the $A$-term contributions to
$B_{\mu}$ is larger than that of the gaugino contributions. If
$\lambda_T>\lambda_D$, the gluino mass is smaller than the value that
GUT relation is valid. Since leading parts in $A$-term contribution is proportional to
gluino mass, the decrease in gluino mass enhances the cancellation
between two classes of contributions. As a result, $B_{\mu}$ becomes
smaller at the electroweak scale and larger $\tan\beta$ is predicted.
Our result is consistent with those in Ref.~\cite{Rattazzi:1996fb}.
 
%%%%%%%%%%%%%%%%%%FIGURE%%%%%%%%%%%%%%%%%%%

%%%%%%%%%%%%%%%%%%%%%%%%%%%%%%%%%%%%%%%%%

%%%%%%%%%%%%%%%%%%%%%%%%%%%%%%%%%%%%%%
\section{Conclusions and discussion} \label{sec:conc}
%%%%%%%%%%%%%%%%%%%%%%%%%%%%%%%%%%%%%%
In this paper we have studied vacuum stability constraints on
left-right mixing of stau's. $\mu\tan\beta$ is bound by quantum
transition to the charge-breaking vacuum.  We have shown lowerbound on
stau mass from the constraint and its corresponding mixing
angle.  Our estimation to vacuum transition rate is based on
semiclassical technique performed in the three-fields space, 
  which give more precise results than previous works.

We have also considered these constraints in the low-energy MGM 
  model. The vacuum stability conditions constrain parameters of
model more severely than experimental limit of stau mass in the case
that messenger triplet is heavier than doublet.  Such split of
couplings in messenger is expected from RG running.

One of issues that we have not handled in this paper is the vacuum
transition in the hot universe. The condition that thermal transition to
charge-breaking vacuum is suppressed enough would also 
restrict $\mu\tan\beta$. In addition,  electroweak symmetric vacuum
should not decay to the charge-breaking vacuum at finite temperature.
To trace electroweak-breaking process properly is 
beyond the scope of this paper and will be studied elsewhere.  

%%%%%%%%%%%%%%%%%%%%%%%%%%%%%%%%%%%%%%%%%%%%
%\begin{acknowledgements}
%%%%%%%%%%%%%%%%%%%%%%%%%%%%%%%%%%%%%%%%%%%%

%%%%%%%%%%%%%%%%%%%%%%%%%%%%%%%%%%%%
\section*{Acknowledgment}
%%%%%%%%%%%%%%%%%%%%%%%%%%%%%%%%%%%%

The work was supported in part by the Grant-in-Aid for the Ministry of
Education, Culture, Sports, Science, and Technology, Government of
Japan, No. 20244037, No. 2054252 and No. 2244021 (J.H.), and also by
the Sasakawa Scientific Research Grant from The Japan Science Society
(S.S.).  The work of J.H. is also supported by the World Premier
International Research Center Initiative (WPI Initiative), MEXT,
Japan.

%%%%%%%%%%%%%%%%%%%%%%%%%%%%%%%%%%%%
\section*{Note Added}
%%%%%%%%%%%%%%%%%%%%%%%%%%%%%%%%%%%%

After this article is published, we found an error in our numerical code. 
All figures and a fitting formula in Eq.~(\ref{fitting}) in this manuscript are replaced from the original ones. The numerical differences between them are at most 15\%. We would like to thank very much M.~Carena, S.~Gori, I.~Low, N.~Shah and C.~Wagner for informing their results to us.

%%%%%%%%%%%%%%%%%%%%%%%%%%%%%%%%%%%%%%%%%%%%
%\end{acknowledgements}
%%%%%%%%%%%%%%%%%%%%%%%%%%%%%%%%%%%%%%%%%%%%

%%%%%%%%%%%%%%%%%%%%%%%%%%%%%%%%%%%%
{}
%%%%%%%%%%%%%%%%%%%%%%%%%%%%%%%%%%%%

\end{document}